\documentclass[conference,a4paper]{APSIPA2020}
\usepackage{multirow,delarray}
\usepackage{subfigure,verbatim}
\usepackage{graphicx}
\usepackage{booktabs}
\usepackage{amsmath}
\usepackage{array}
\usepackage{diagbox, makecell}
\usepackage[psamsfonts]{amssymb}
\usepackage{amsxtra}
\usepackage{threeparttable}

\begin{document}

\title{A Time-domain Monaural Speech Enhancement with Feedback Learning}

\author{%
	\authorblockN{%
		Andong Li\authorrefmark{1}\authorrefmark{2} Chengshi Zheng\authorrefmark{1}\authorrefmark{2} Linjuan Cheng\authorrefmark{1}\authorrefmark{2} Renhua Peng\authorrefmark{1}\authorrefmark{2} and Xiaodong Li\authorrefmark{1}\authorrefmark{2}
	}
	\authorblockA{%
		\authorrefmark{1}
		Key Laboratory of Noise and Vibration Research, Institute of Acoustics, Chinese Academy of Sciences, Beijing, China\\
		}
	\authorblockA{%
		\authorrefmark{2}
		University of Chinese Academy of Sciences, Beijing, China\\
	E-mail: \{liandong, cszheng, chenglinjuan, pengrenhua, lxd\} @mail.ioa.ac.cn}
}

\maketitle
\thispagestyle{empty}

\begin{abstract}
  In this paper, we propose a type of neural network with feedback learning in the time domain called FTNet for monaural speech enhancement, where the proposed network consists of three principal components. The first part is called stage recurrent neural network, which is introduced to effectively aggregate the deep feature dependencies across different stages with a memory mechanism and also remove the interference stage by stage. The second part is the convolutional auto-encoder. The third part consists of a series of concatenated gated linear units, which are capable of facilitating the information flow and gradually increasing the receptive fields. Feedback learning is adopted to improve the parameter efficiency and therefore, the number of trainable parameters is effectively reduced without sacrificing its performance. Numerous experiments are conducted on TIMIT corpus and experimental results demonstrate that the proposed network can achieve consistently better performance in terms of both PESQ and STOI scores than two state-of-the-art time domain-based baselines in different conditions. 
\end{abstract}

\section{Introduction}
\label{sec:intro}
  Speech is often inevitably degraded by background interference in real environments, which may significantly reduce the performance of automatic speech recognition (ASR), speech communication system and hearing aids. Monaural speech enhancement is dedicated to effectively extracting underlying target speech from its degraded version when only one measurement is available~{\cite{loizou2013speech}}. There are many well-known unsupervised signal-processing-based approaches, such as spectral subtraction~{\cite{boll1979suppression}}, Wiener filtering~{\cite{hu2013cepstrum}} and statistical-based methods~{\cite{jensen1995reduction}}.
  
  Recent advances in deep neural networks (DNNs) have facilitated the rapid development of speech enhancement research, and a great number of DNN models have been proposed to tackle the nonlinear mapping problem from the noisy speech to the clean speech (see~{\cite{wang2018supervised, xu2015regression}} and references therein). A typical DNN-based speech enhancement framework extracts time-frequency (T-F) features of the noisy speech and calculates some T-F representation targets of the clean speech. A model is then trained to establish the complicated mapping from the input features to the output targets with some supervised methods. Training targets can be categorized into two types, where one is the masking-based~{\cite{wang2014training}} and the other is the spectral mapping-based~{\cite{xu2015regression, tan2018convolutional}}.
  
  Different from the research line in the T-F domain~{\cite{xu2015regression, wang2014training, tan2018convolutional}}, a multitude of approaches based on time domain has emerged more recently~{\cite{yiluo2019, pascual2017segan, rethage2018wavenet, pandey2019new, abdulbaqi2019rhr}}. Compared with T-F domain based methods, the major advantage of time domain approaches is that the phase estimation problem can be mitigated, which is helpful for speech quality~{\cite{paliwal2011importance}}. Pandey et al.~{\cite{pandey2019new}} took the U-Net with fully convolutional networks (FCNs) to directly model the waveform and utilized the domain knowledge from the time domain to the frequency domain to optimize the loss, which was significant for spectral detail recovery. Pascual et al. first applied the generative adversarial network (GAN) into the speech enhancement task in the time domain, where the generator was trained to produce a cleaner waveform whilst the discriminator was enforced to distinguish between the fake and clean versions. Luo et al.~{\cite{yiluo2019}} utilized a learned encoder and decoder to project the speech waveform into a latent space, and superior performance was observed than short-time Fourier transform (STFT) based approaches in the speech separation task.

  
  Despite the success of time-domain based approaches in the speech enhancement task~{\cite{pascual2017segan, rethage2018wavenet, pandey2019new, abdulbaqi2019rhr}}, these processing systems require a large number of trainable parameters, which may increase the computational complexity for practical applications. More recently, progressive learning (PL) has been applied in various tasks like single image deraining~{\cite{ren2019progressive}} and speech enhancement~{\cite{gao2018densely}}, where the whole mapping procedure is decomposed into multiple stages. In our preliminary work, we propose a PL-based convolutional recurrent network (PL-CRN)~{\cite{Li2019ConvolutionalRN}}, where the noise components are gradually attenuated with a light-weight convolutional recurrent network (CRN) in each stage. We attribute the success of PL to the accumulation of prior information with the increase of the stages, i.e., all the outputs in the previous stages actually serve as the prior information to facilitate the execution of subsequent stages. Motivated by these studies, we propose a novel time-domain-based network with a feedback mechanism called FTNet, which needs much fewer trainable parameters. It works by recursively incorporating the estimated output from the last stage along with the original noisy feature back to the network, where each temporary output can be regarded as a type of state among different stages and thus trained with a recurrent approach. By doing so, the feature dependencies across different stages can be fully exploited and the output estimation can be refined stage by stage. 

  The remainder of this paper is structured as follows. Section~{\ref{sec:network-module}} formulates the problem and briefly introduces the principal modules of the network. The proposed architecture is described in Section~{\ref{sec:proposed-architecture}}. Section~{\ref{sec:experiments}} presents the experimental settings. Experimental results and analysis are given in Section~{\ref{sec:results-and-analysis}}. Some conclusions are drawn in Section~{\ref{sec:conclusions}}.

\section{Network module}
\label{sec:network-module}
  In the time domain, a mixture signal is usually formulated as $x(k) = s(k) + d(k)$, where $k$ denotes the time index, $s(k)$, $d(k)$, and $x(k)$ are the clean speech, the noise, and the noisy speech, respectively. The network aims to estimate the time-domain clean speech. For notation convenience, we denote the frame vector of the noisy signal, estimation in $l$th stage, and the final output in the time domain as $\mathbf{x}\in \mathbb{R}^{K}$, $\mathbf{\tilde{s}}^{l}\in \mathbb{R}^{K}$, $\mathbf{\tilde{s}}\in \mathbb{R}^{K}$, respectively, where $K$ is the frame length and $l$ is the stage index. The proposed architecture is in essence a type of multi-stage network, where the output speech is estimated and refined stage by stage. Assuming the number of training stages is denoted as Q, in each stage, the estimated output from the last stage and the original noisy input are combined and sent back to the network. For the $l$th stage, the mapping process can be formulated as:
\begin{equation}
\setlength{\abovedisplayskip}{2pt}
\setlength{\belowdisplayskip}{2pt}
\label{eqn:time-domain}
\mathbf{\tilde{s}}^{l} = g_{\theta}(\mathbf{x},\mathbf{\tilde{s}}^{l-1}),
\end{equation}
where $g_{\theta}(.)$ represents the network function. As seen from Eq.~{\ref{eqn:time-domain}}, both the estimation from the last stage and original noisy input are connected to update the estimation in the current stage.

\subsection{Stage recurrent neural network}
\label{SRNN}

\begin{figure}[t]
	\begin{center}
		\includegraphics[width=\columnwidth]{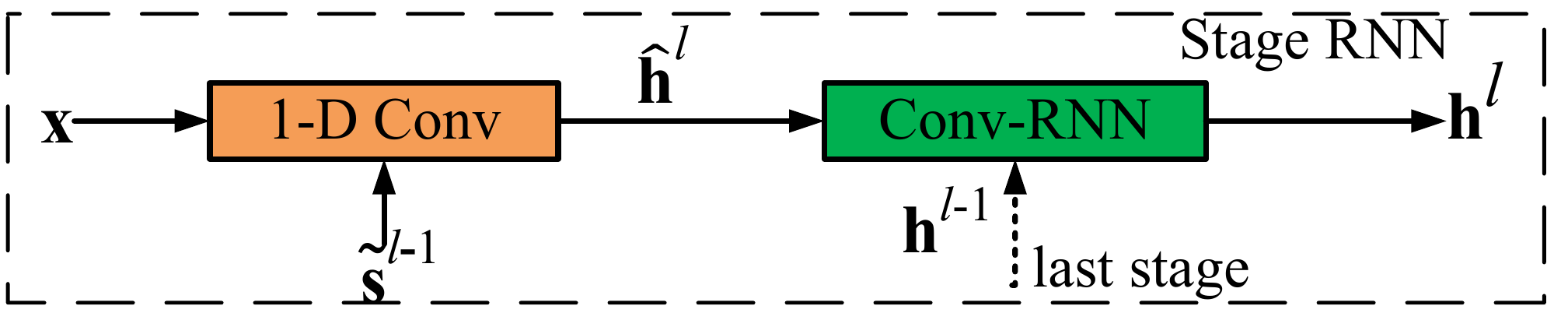}
	\end{center}
	\caption{The internal detail of SRNN module. It includes a 1-D Conv block and a Conv-RNN block. The module is operated with double input and single output (DISO).}
	\label{fig:stage_rnn}
\end{figure}
	Theoretically, the learning process from the noisy feature to the clean target can be viewed as a type of sequence learning, where each state represents the intermediate output in one stage. To this end, we propose a type of recurrent convolutional structure named stage recurrent neural network (SRNN) to explore the time dependencies of different stages in this study. As a result, the network can be trained following a recurrent learning paradigm. As shown in Fig.~\ref{fig:stage_rnn}, SRNN contains two parts, namely 1-D Conv block and convolutional-RNN (Conv-RNN). Assuming the inputs are $\mathbf{x}$ and $\mathbf{\tilde{s}}^{l-1} $, and the output of the 1-D Conv block is denoted as $\mathbf{\hat h}^{l}$. Then $\mathbf{\hat h}^{l}$ along with the hidden state vector from the last stage $\mathbf{h}^{l-1}$ is sent to Conv-RNN to obtain a updated hidden state, i.e., $\mathbf{h}^{l}$. As a result, the inference of $\mathbf{h}^{l}$ can be formulated as
\begin{gather}
\setlength{\abovedisplayskip}{2pt}
\setlength{\belowdisplayskip}{2pt}
\label{eqn:stage_rnn}
\mathbf{\hat h}^{l} = f_{conv}(\mathbf{x},    \mathbf{\tilde{s}}^{l-1}),\\
\mathbf{h}^{l} = f_{conv\_rnn}(\mathbf{\hat h}^{l}, \mathbf{h}^{l-1}),
\end{gather}
where $f_{conv}(\cdot)$ and $f_{conv\_rnn}(\cdot)$ represent the functions of 1-D Conv block and Conv-RNN block, respectively.

In this study, ConvGRU~\cite{ballas2015delving} is adopted as the unit for Conv-RNN, given as follows:
\begin{gather}
\label{eqn:conv_gru}
\mathbf{z}^{l} = \sigma\left(\mathbf{W}^{l}_{z}\circledast\mathbf{\hat h}^{l} + \mathbf{U}^{l}_{z}\circledast\mathbf{h}^{l-1} \right),\\
\mathbf{r}^{l} = \sigma\left(\mathbf{W}^{l}_{r}\circledast\mathbf{\hat h}^{l} + \mathbf{U}^{l}_{r}\circledast\mathbf{h}^{l-1} \right),\\
\mathbf{n}^{l} = \tanh\left(\mathbf{W}^{l}_{n}\circledast\mathbf{\hat h}^{l} + \mathbf{U}^{l}_{n}\circledast\left(\mathbf{r}^{l}\odot\mathbf{h}^{l-1}\right) \right),\\
\mathbf{h}^{l} = \left(\mathbf{1} - \mathbf{z}^{l}\right)\odot\mathbf{\hat h}^{l} + \mathbf{z}^{l}\odot\mathbf{n}^{l},
\end{gather}
where $\sigma(\cdot)$ and $\tanh(\cdot)$, respectively, denote the sigmoid and the tanh activation functions. $\mathbf{W}$ and $\mathbf{U}$ refer to the weight matrices of the cell. $\circledast$ represents the convolutional operator and $\odot$ is the element-wise multiplication. Note that all the biases are neglected for notation simplicity.
\subsection{Gated linear unit}
\label{sec:glu}

\begin{figure*}[t]
	\centering
	\centerline{\includegraphics[width=2\columnwidth]{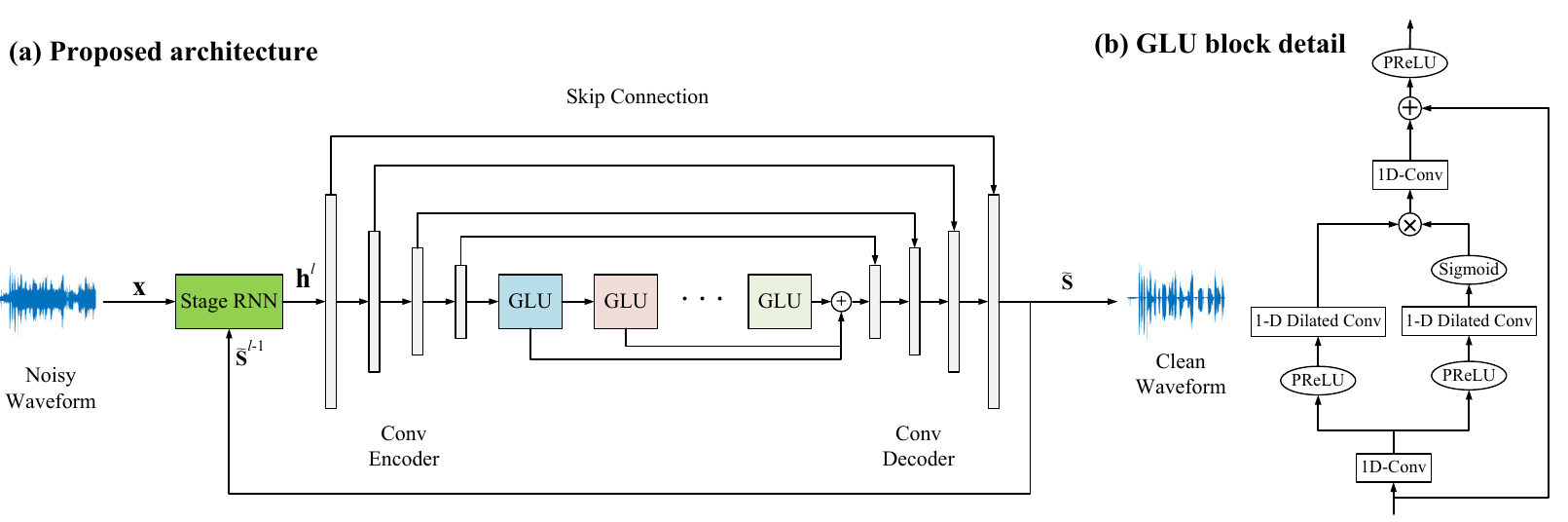}}
	\caption{The framework of proposed network FTNet with feedback learning. (a) The overview of FTNet. $\mathbf{x}$, $\tilde{\mathbf{s}}^{l-1}$, $\tilde{\mathbf{h}}^{l}$ and $\tilde{\mathbf{s}}$ denote the input feature, the estimation output in stage $l-1$, the state in stage $l$ and the final estimation output, respectively. (b) The detail of GLU adopted in this study, where PReLU is adopted and the kernel size is set to 11.}
	\label{fig:proposed-architecture}
\end{figure*}

Gated convolutional layer is first introduced in~\cite{van2016conditional} to model complicated interactions in the form of a gating mechanism which is beneficial to performance and its modified version named GLU is utilized in~\cite{tan2018gated} by replacing the $\tanh$ nonlinearity with a linear unit and residual learning is also incorporated to mitigate gradient vanishing problem when learning deep features~\cite{resenet16}. In this study, we stack multiple GLU modules to explore the sequence correlations among neighboring points. As shown in Fig.~\ref{fig:proposed-architecture}-(b), two additional branches are introduced compared with the conventional CNN block, where one is the gated operation that is controlled with the sigmoid function to adjust the information flow percentage and the other is residual connection. Dilated convolution is applied to increase the receptive field, which is beneficial to capture more sequence correlations. We use parametric ReLU (PReLU)~{\cite{he2015delving}} as the activation function and the kernel size is set to 11 herein.

{
	\renewcommand\arraystretch{0.8}
	\setlength{\tabcolsep}{5pt}
	\begin{table}[t]
		\tiny
		\footnotesize
		\caption{Detailed parameter setup of the proposed architecture.}
		\vspace{-0.4cm}
		\begin{center}
			\begin{tabular}{c|c|c|c}
				\hline
				layer name & input size & hyperparameters & output size\\
				\hline
				conv1d\_1 & 2 $\times$ 2048 & (11, 2, 16) & 16 $\times$ 1024\\
				\hline
				conv\_rnn & 16 $\times$ 1024 & (11, 1, 16) & 16 $\times$ 1024\\
				\hline
				conv1d\_2 & 16 $\times$ 1024 & (11, 1, 16) & 16 $\times$ 1024\\
				\hline
				conv1d\_3 & 16 $\times$ 1024 & (11, 2, 32) & 32 $\times$ 512\\
				\hline
				conv1d\_4 & 32 $\times$ 512 & (11, 2, 64) & 64 $\times$ 256\\
				\hline
				conv1d\_5 & 64 $\times$ 256 & (11, 2, 128)& 128 $\times$ 128\\
				\hline
				GLUs & 128 $\times$ 128 & $
				\begin{array}{c}
				\begin{array}({c})
				1, 1, 64 \\ 11, \mathbf{1}, 64 \\1, 1, 128
				\end{array}\\
				\begin{array}({c})
				1, 1, 64 \\ 11, \mathbf{2}, 64 \\1, 1, 128
				\end{array}\\
				\begin{array}({c})
				1, 1, 64 \\ 11, \mathbf{4}, 64 \\1, 1, 128
				\end{array}\\
				\begin{array}({c})
				1, 1, 64 \\ 11, \mathbf{8}, 64 \\1, 1, 128
				\end{array}\\
				\begin{array}({c})
				1, 1, 64 \\ 11, \mathbf{16}, 64 \\1, 1, 128
				\end{array}\\
				\begin{array}({c})
				1, 1, 64 \\ 11, \mathbf{32}, 64 \\1, 1, 128
				\end{array}
				\end{array}
				$ & 128 $\times$ 128\\
				\hline
				skip\_1  & 128 $\times$ 128 & - & 256 $\times$ 128\\
				\hline
				deconv1d\_1 & 256 $\times$ 128 & (11, 2, 64)& 64 $\times$ 256\\
				\hline
				skip\_2  & 64 $\times$ 256 & - & 128 $\times$ 256\\
				\hline
				deconv1d\_2 & 128 $\times$ 256 & (11, 2, 32) & 32 $\times$ 512\\
				\hline
				skip\_3  & 32 $\times$ 512 & - & 64 $\times$ 512\\
				\hline
				deconv1d\_3 & 64 $\times$ 512 & (11, 2, 16) & 16 $\times$ 1024\\
				\hline
				skip\_4  & 16 $\times$ 1024 & - & 32 $\times$ 1024\\
				\hline
				deconv1d\_4 & 32 $\times$ 1024 & (11, 2, 1) & 1 $\times$ 2048\\
				\hline
			\end{tabular}
			\label{tab1}
		\end{center}
	\end{table}
} 

\section{Proposed architecture}
\label{sec:proposed-architecture}
The architecture of FTNet is illustrated in Fig.~\ref{fig:proposed-architecture}-(a), which includes three parts, namely SRNN, convolutional auto-encoder (CAE)~{\cite{badrinarayanan2015segnet}} and a series of GLUs. SRNN consists of a 1-D Conv block and a ConvRNN block. 1-D Conv takes the concatenation of both noisy speech vector and the output estimation vector from the last stage along the channel axis. Therefore, the size of network input is $\left(2, K \right)$, where $2$ refers to channels. After SRNN, the output is sent to the subsequent modules. CAE consists of the convolutional encoder and the decoder. The encoder consists of four 1-D Conv blocks, which compresses and establishes the deep representation of the features by halving the feature length with strided operation while consecutively doubling the channels. The decoder is the symmetric representation compared with the encoder, where the length of the feature is successively expanded through a number of deconvolutional layers~{\cite{noh2015learning}}. Both encoder and decoder adopt PReLU as the activation nonlinearity except the output layer, where $\tanh$ is used to normalize the value range into $\left[ -1, 1 \right]$. Additionally, skip connections are adopted to connect each encoding layer to its homologous decoding layer, which compensates for the feature loss during the encoding process. To model the time correlations, six concatenated GLUs are inserted between the encoder and decoder, where the dilated rates are $\left( 1, 2, 4, 8, 16, 32 \right)$.

When the estimation output of the $l$th stage is obtained, i.e., $\mathbf{\tilde{s}}^{l}$, it is fed back and concatenated with the noisy input $\mathbf{x}$ along channel axis to execute the next stage. Here we only impose supervision on the final output $\mathbf{\tilde s}$, which is consistent with the setting in~{\cite{ren2019progressive}}.

A more detailed parameter configuration of the proposed network is summarized in Table~\ref{tab1}, where the input and output sizes of 2-D tensor representation are specified with $\left( Channels \times Framesize \right)$ format. The hyperparameters of the layers except GLUs are specified with $\left( KernelSize, Strided, Channels \right)$ format. The hyperparameters of GLUs are specified with $\left( KernelSize, DilatedRate, Channels \right)$ format. Bold numbers refer to the dilated rate.

\section{Experiments}
\label{sec:experiments}
\subsection{Datasets}

\label{sec:datasets}
Experiments are conducted on TIMIT corpus~{\cite{garofolo1993darpa}}, which includes 630 speakers of eight major dialects of American English with each reading ten utterances. 1000, 200 and 100 clean utterances are randomly selected for training, validation and testing, respectively. Training and validation dataset are mixed under different SNR levels ranging from -5{\rm dB} to 10{\rm dB} with the interval 1dB while the testing datasets are mixed under -5{\rm dB} and -2{\rm dB} conditions. For training and validation, we use 130 types of noises, including 115 types used in~{\cite{Li2019ConvolutionalRN}}, 9 types from~{\cite{duan2012speech}}, 3 types from NOISEX92~{\cite{varga1993assessment}} and 3 common environmental noise, i.e. aircraft, bus and cafeteria. Another 5 types of noises from NOISEX92, including babble, f16, factory2, m109 and white, are chosen to test the network generalization capacity.

Various noises are first concatenated into a long vector. During each mixed process, the cutting point is randomly generated, which is subsequently mixed with a clean utterance under one SNR condition. As a result, totally 10,000, 2000 and 400 noisy-clean utterance pairs are created for training, validation, and testing, respectively. 

\subsection{Baselines}
\label{baselines}
In this study, two advanced time-domain-based networks are selected as the baselines, namely AECNN~{\cite{pandey2019new}} and RHR-Net~{\cite{abdulbaqi2019rhr}}. AECNN is a typical 1-D Conv based auto-encoder architecture with a large number of trainable parameters. The number of channels in consecutive layers are \{64, 64, 64, 128, 128, 128, 256,  256, 256, 512, 512, 256, 256, 256, 128, 128, 128, 1\}, with 11 and PReLU being the filter size and activation nonlinearity, respectively. RHR-Net has also the form of auto-encoder framework except all the convolutional layers are replaced by bidirectional GRU (BiGRU). In addition, direct skip connections are replaced by PReLU based residual connections. It achieves state-of-the-art metric performance among several advanced speech enhancement models with limited trainable parameters (see~\cite{abdulbaqi2019rhr}). The number of units per layer are \{1, 32, 64, 128, 256, 128, 64, 32, 1\} and three residual skip connections are introduced. Note that the last layer is a single-directional GRU to output the enhanced signal.

\renewcommand\arraystretch{0.8}
\begin{table}[t]
	\caption{Experimental results under seen noise conditions for PESQ and STOI. \textbf{BOLD} indicates the best result for each case. The number of stages Q are set to 3, 4 and 5 for model comparisons.}
	\centering
	\scriptsize
	\resizebox{0.46\textwidth}{!}{
		\begin{tabular}{ccccccc}
			\toprule
			\multicolumn{1}{c}{Metrics} & \multicolumn{3}{c}{PESQ}  & \multicolumn{3}{c}{STOI (in \%)} \\
			\midrule
			\multicolumn{1}{c}{SNR} & -5\rm{dB}  &-2\rm{dB} &Avg.
			& -5\rm{dB}  &-2\rm{dB} &Avg.\\
			\midrule
			\multicolumn{1}{c}{Noisy}  &1.47 &1.66 &1.57 &63.03 &68.20 &65.62 \\
			AECNN &2.25 &2.49 &2.37 &82.70 &87.51 &85.11 \\
			RHR-Net &2.32 &2.55 &2.44 &83.13 &87.90 &85.51 \\
			FTNet (Q = 3) &2.36 &2.59 &\textbf{2.48} &83.18 &87.92 &85.55 \\
			FTNet (Q = 4) &2.35 &2.59 &2.47 &83.75 &88.39 &86.07 \\
			FTNet (Q = 5) &\textbf{2.37} &\textbf{2.60} &\textbf{2.48} &\textbf{84.03} &\textbf{88.54} &\textbf{86.28} \\
			\bottomrule
	\end{tabular}}
	\label{tbl:seen-results1}
	\vspace{-0cm}
\end{table}

\renewcommand\arraystretch{0.8}
\begin{table}[t]
	\caption{Experimental results under unseen noise conditions for PESQ and STOI. \textbf{BOLD} indicates the best result for each case. The number of stages Q are set to 3, 4 and 5 for model comparisons.}
	\centering
	\scriptsize
	\resizebox{0.46\textwidth}{!}{
		
		\begin{tabular}{ccccccc}
			\toprule
			\multicolumn{1}{c}{Metrics} & \multicolumn{3}{c}{PESQ}  & \multicolumn{3}{c}{STOI (in \%)} \\
			\midrule
			\multicolumn{1}{c}{SNR} & -5\rm{dB}  &-2\rm{dB} &Avg.
			& -5\rm{dB}  &-2\rm{dB} &Avg.\\
			\midrule
			\multicolumn{1}{c}{Noisy}  &1.44 &1.67 &1.56 &59.64 &67.45 &63.55 \\
			AECNN &1.88 &2.20 &2.04 &77.37 &85.10 &81.24 \\
			RHR-Net &2.06 &2.35 &2.21 &78.13 &85.82 &81.98 \\
			FTNet (Q = 3) &\textbf{2.10} &\textbf{2.37} &\textbf{2.23} &78.59 &85.68 &82.13 \\
			FTNet (Q = 4) &2.06 &2.35 &2.21 &79.31 &86.20 &82.76 \\
			FTNet (Q = 5) &2.09 &2.35 &2.22 &\textbf{79.48} &\textbf{86.54} &\textbf{83.01} \\
			\bottomrule
	\end{tabular}}
	\label{tbl:unseen-results1}
\end{table}
\subsection{Experimental settings}
\label{sec:experimental}
We sample all the utterances at 16kHz. Each frame has a size of 2048 samples (128 ms) with 256 samples (16 ms) offset between adjacent frames. All the models are trained with mean absolute error (MAE) criterion, optimized by Adam algorithm~{\cite{kingma2014adam}}. The learning rate is initialized at 0.0002. We halve the learning rate only if consecutive three validation loss increment arises and the training process is early-stopped only if ten validation loss increment happens. We train all the models for 50 epochs. Within each epoch, the minibatch is set to 2 at the utterance level, where all the utterances are randomly chunked to 4 seconds if they exceed 4 seconds and zero-padded on the contrary. 

\section{Results and analysis}
\label{sec:results-and-analysis}
We evaluate the performance of different models in terms of perceptual evaluation of speech quality (PESQ)~{\cite{rix2001perceptual}} and short-time objective intelligibility (STOI)~{\cite{taal2011algorithm}}.

\subsection{Objective results comparison}
The objective results are presented in Tables~{\ref{tbl:seen-results1}} and~{\ref{tbl:unseen-results1}}. One can observe the following phenomena. Firstly, all the models significantly improve the scores in terms of PESQ and STOI for both seen and unseen cases, whilst the proposed FTNet achieves the best performance among the three models. For example, for seen cases, when Q $= 3$, FTNet improves PESQ by 0.11 and 0.04, and improves STOI by 0.44\% and 0.04\% over AECNN and RHR-Net, respectively. This is because the memory mechanism is utilized to refine the network with a stage-wise manner and improve the parameter efficiency. A similar tendency is also observed for unseen cases. Secondly, when comparing between two baselines, RHR-Net obtains consistently better performance than AECNN. This is because BiGRU is adopted as the basic component for both encoding and decoding process, which facilitates better temporal capture capability for long sequences than 1-D Conv, whose performance is limited by kernel size and dilation rate. This can also partly explain the limited advantages of FTNet over RHR-Net. 

\begin{figure}[t]
	\centering
	\subfigure[]{
		\includegraphics[width= 38mm]{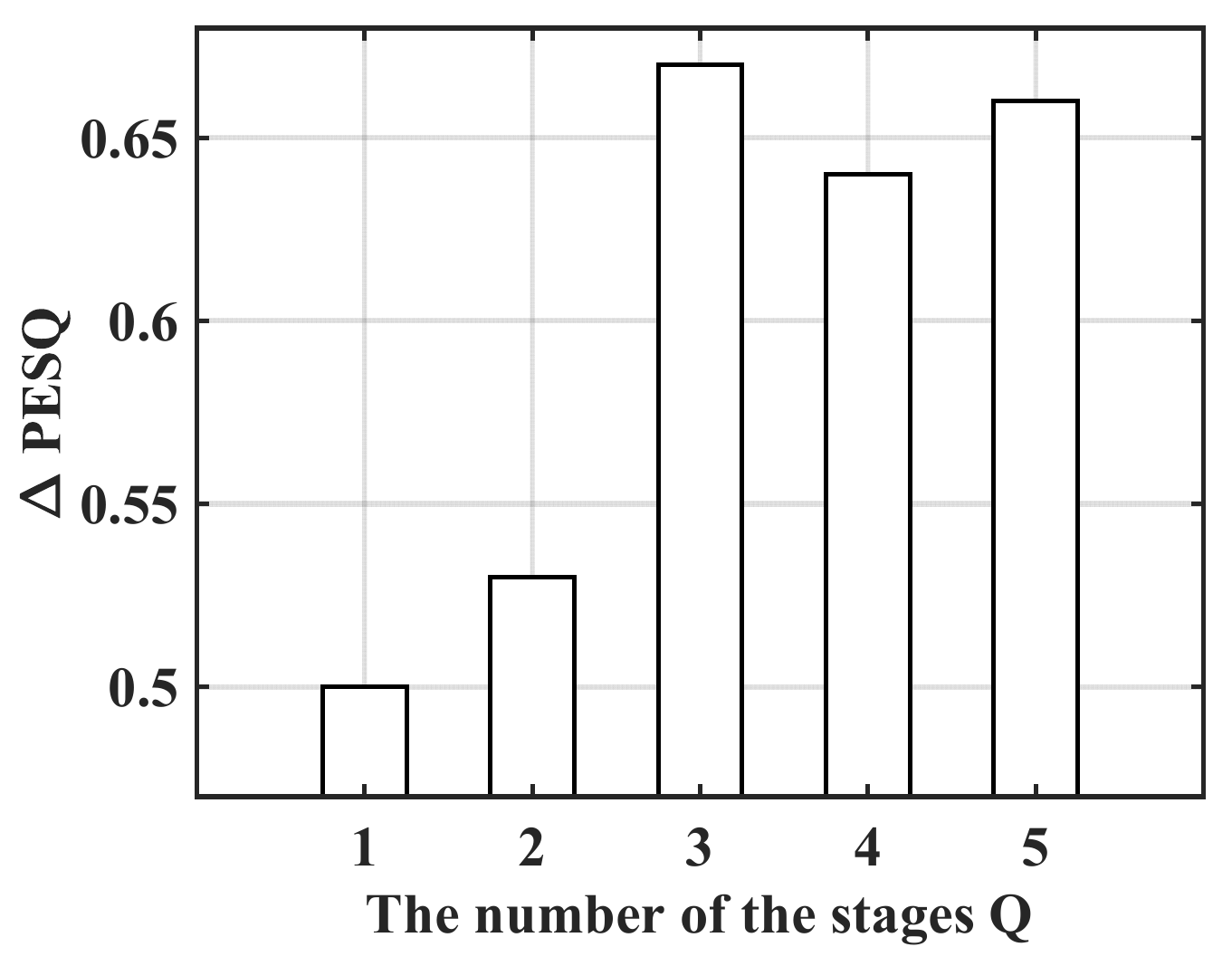}	
	}
	\subfigure[]{
		\includegraphics[width= 38mm]{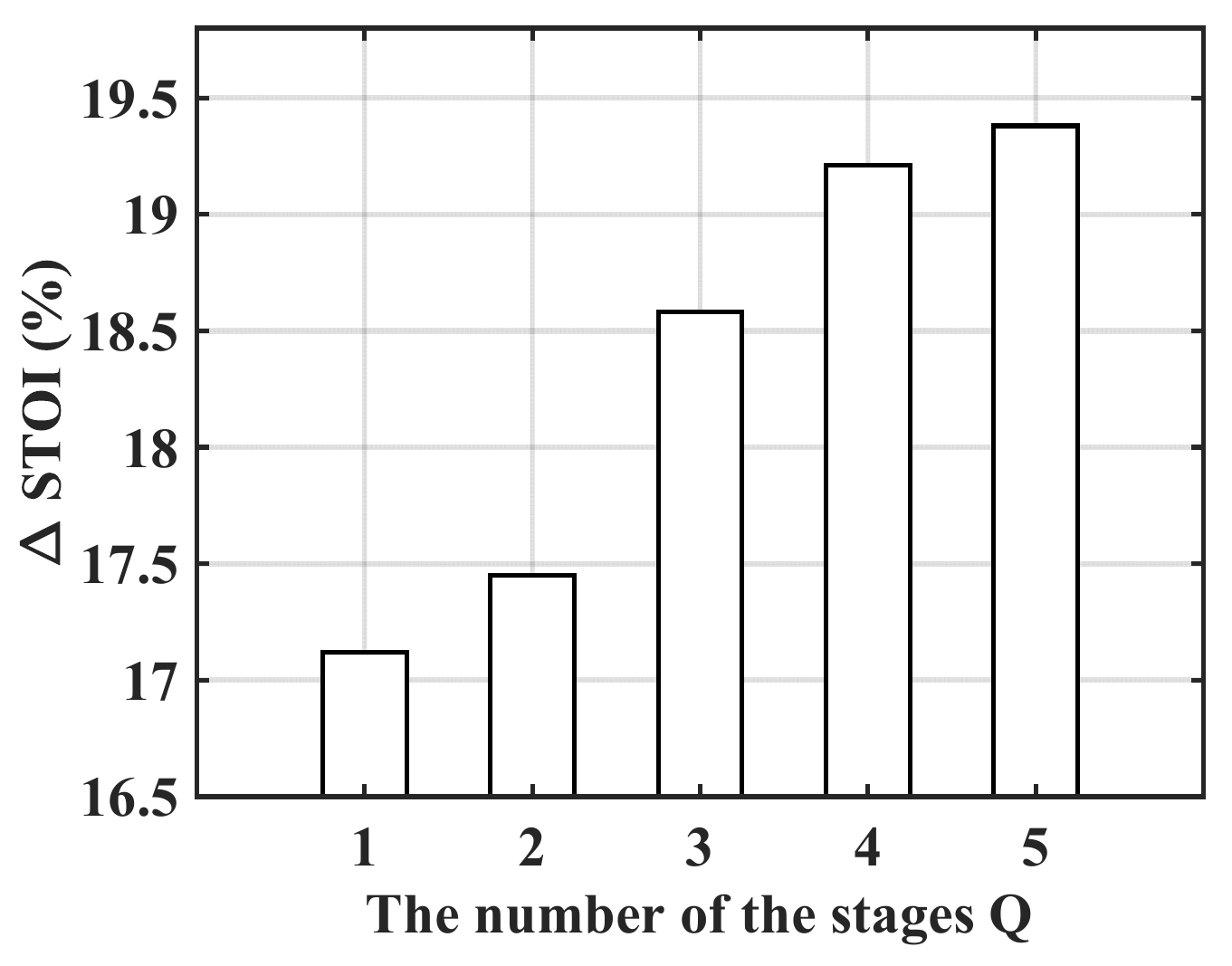}	
	}
	\caption{PESQ and STOI improvements with the increase of the number of the stages Q. The values are averaged over unseen dataset. Here five values are explored, i.e., Q $= 1, 2, 3, 4, 5$. }
	\label{fig:result-stage}
\end{figure}

\subsection{The influence of stage number Q}
In this study, we explore the influence of the number of the stages Q, and it takes the values from 1 to 5. Note that Q $= 1$ means that only one stage is applied and no memory mechanism is adopted to bridge the relationship between neighboring stages. The metric improvements are given in Fig.~{\ref{fig:result-stage}}.  One can observe the following phenomena. Firstly, when Q $\leq 3$, both PESQ and STOI scores are consistently improved with the increase of Q, indicating that both metrics can be effectively refined with feedback learning. Nonetheless, when Q takes from 3 to 5, PESQ falls into saturation even slightly attenuation while STOI is further improved. This is because MAE is adopted as the loss criterion, whose optimization target is inconsistent with the objective evaluation criterion and can not further refine both metrics simultaneously~{\cite{fu2018end}}. This phenomenon reveals that further optimization of MAE can improve STOI but may slightly reduce PESQ. 

\begin{figure}[t]
	\centering
	\subfigure[]{
		\includegraphics[width= 0.45\columnwidth]{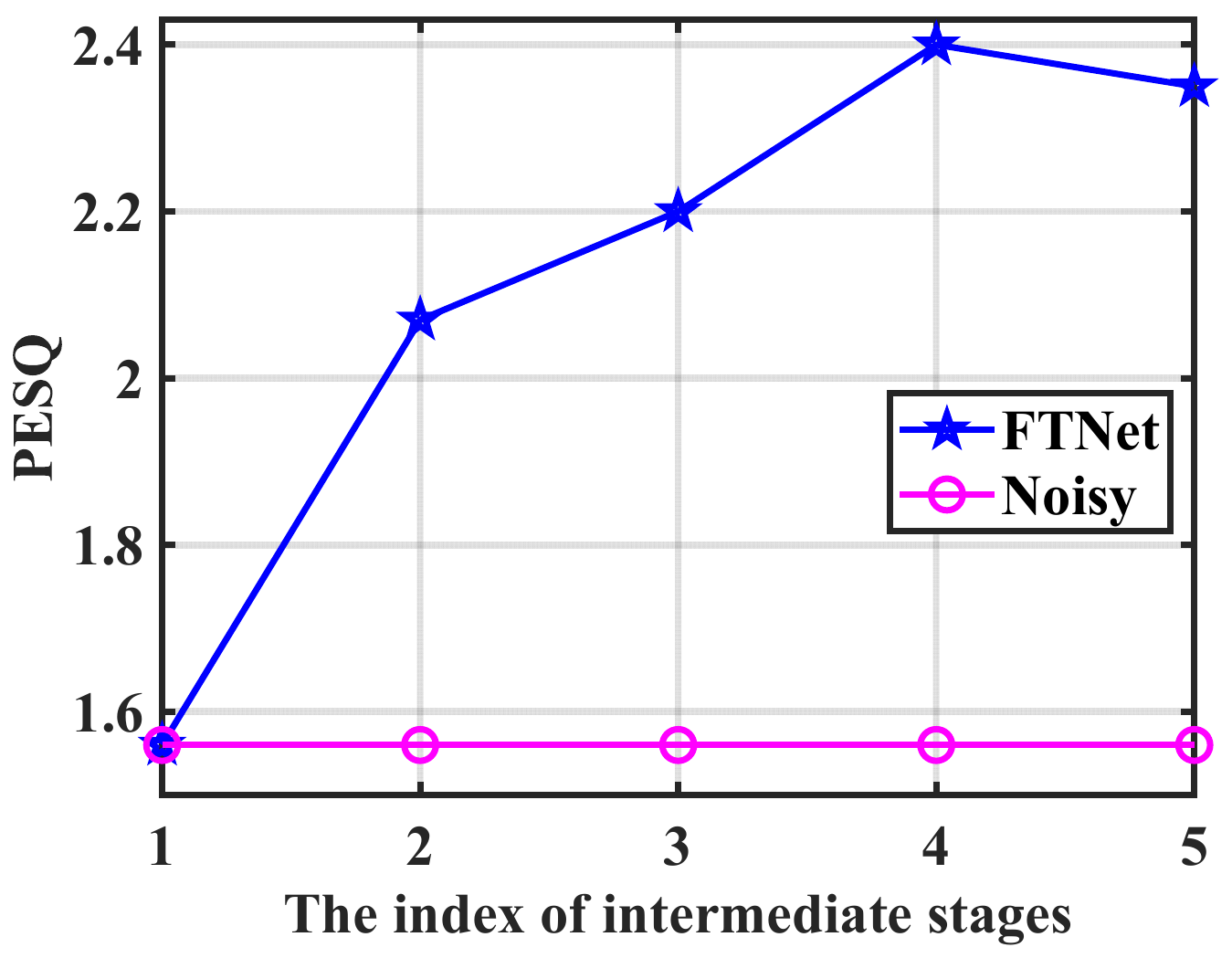}	
	}
	\subfigure[]{
		\includegraphics[width= 0.45\columnwidth]{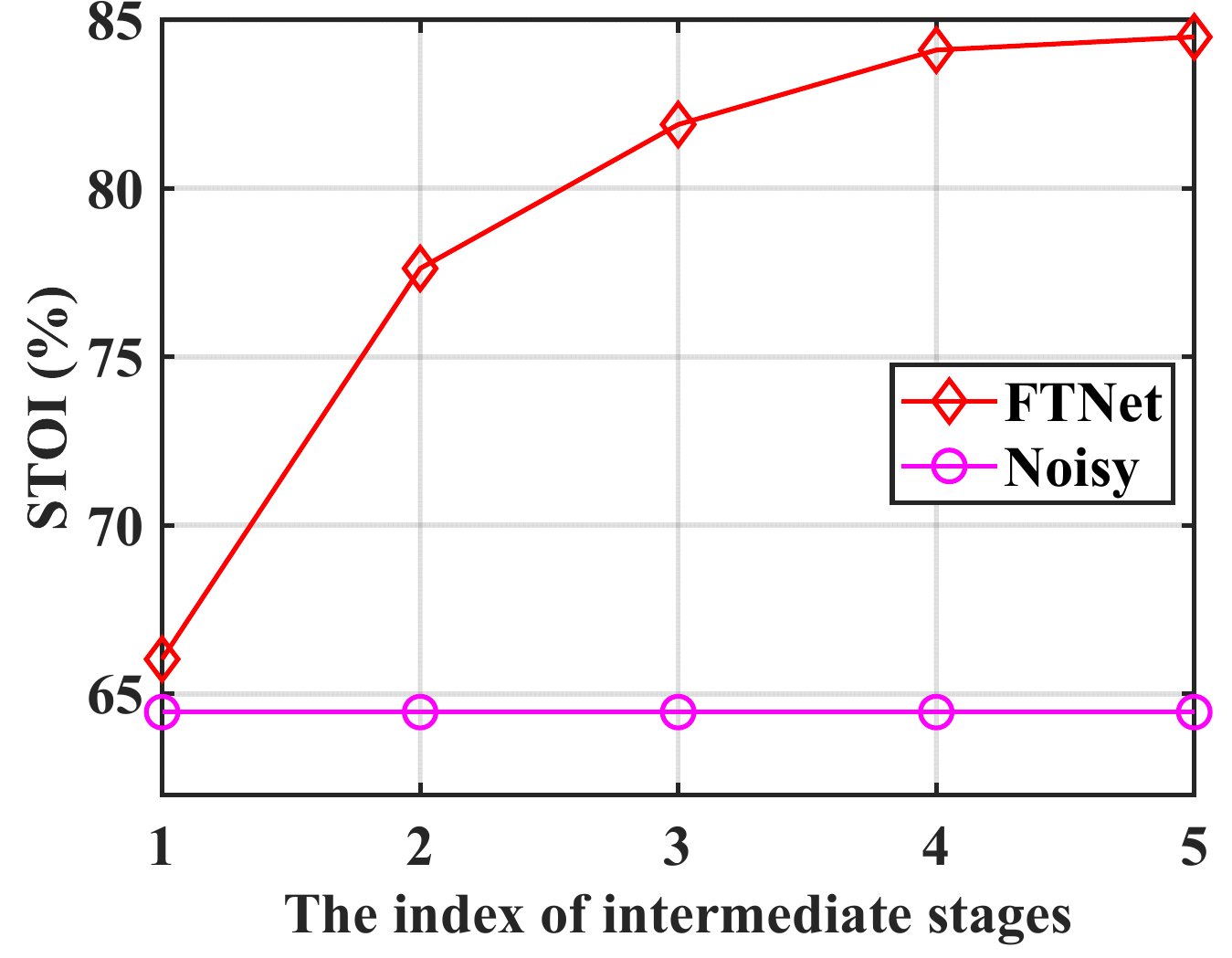}	
	}
	\caption{The metric scores in terms of PESQ and STOI for different intermediate stages given Q $= 5$. The results are averaged over both seen and unseen conditions. Noisy scores are also presented for comparison.}
	\label{fig:results-inter}
\end{figure}

\begin{figure}[t]
	\centering
	\includegraphics[width= 0.9\columnwidth]{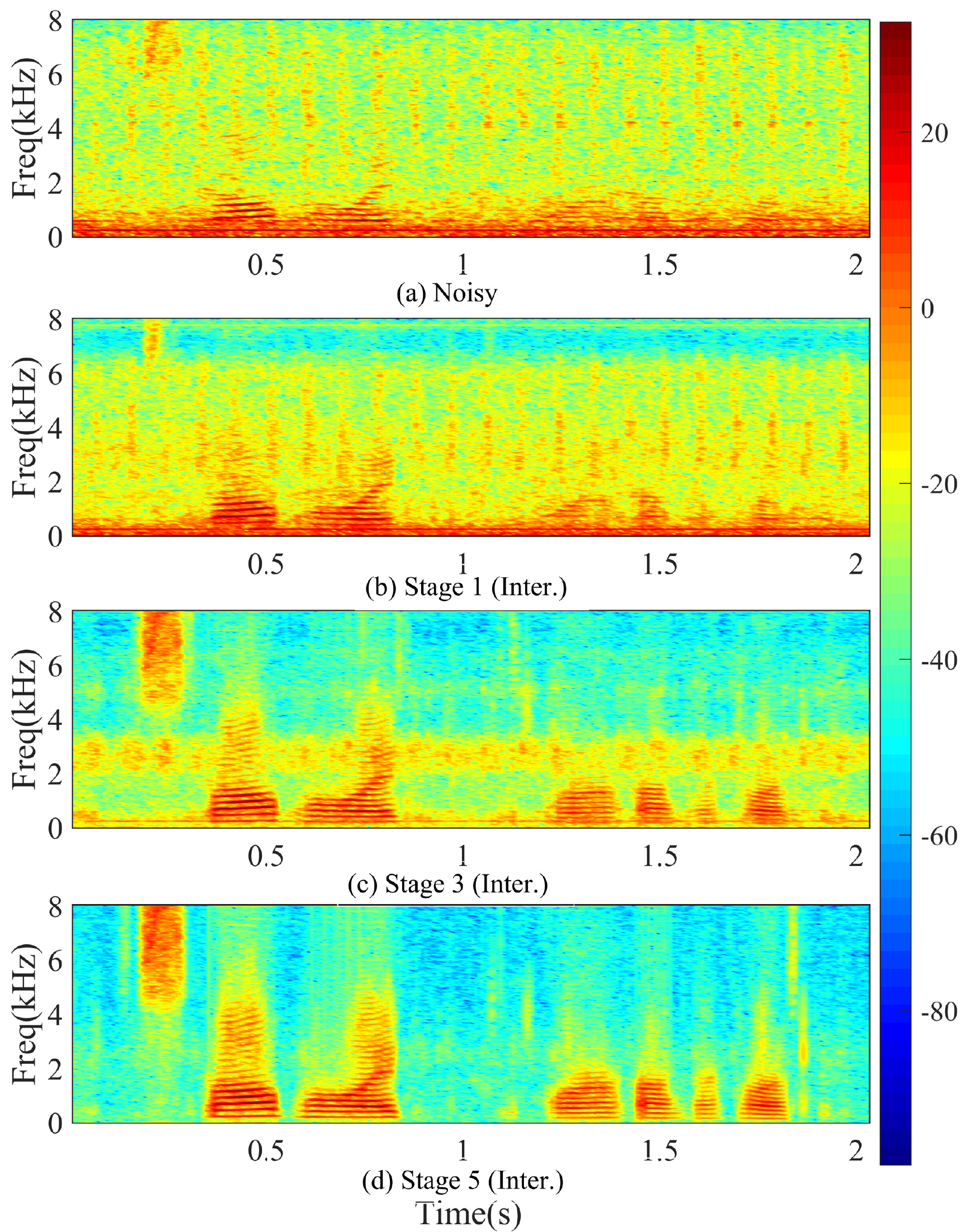}	
	
	\caption{Spectral visualization for different intermediate stages given Q $= 5$. (a) Noisy spectrogram under -5dB, PESQ=0.98. (b) Enhanced spectrogram in the first stage, PESQ=1.06. (c) Enhanced spectrogram in the third stage, PESQ=1.61. (d) Enhanced spectrogram in the fifth stage, PESQ=1.83.}
	\label{fig:mecha-illu}
\end{figure}

\begin{figure}[t]
	\centering
	\includegraphics[width= 0.9\columnwidth]{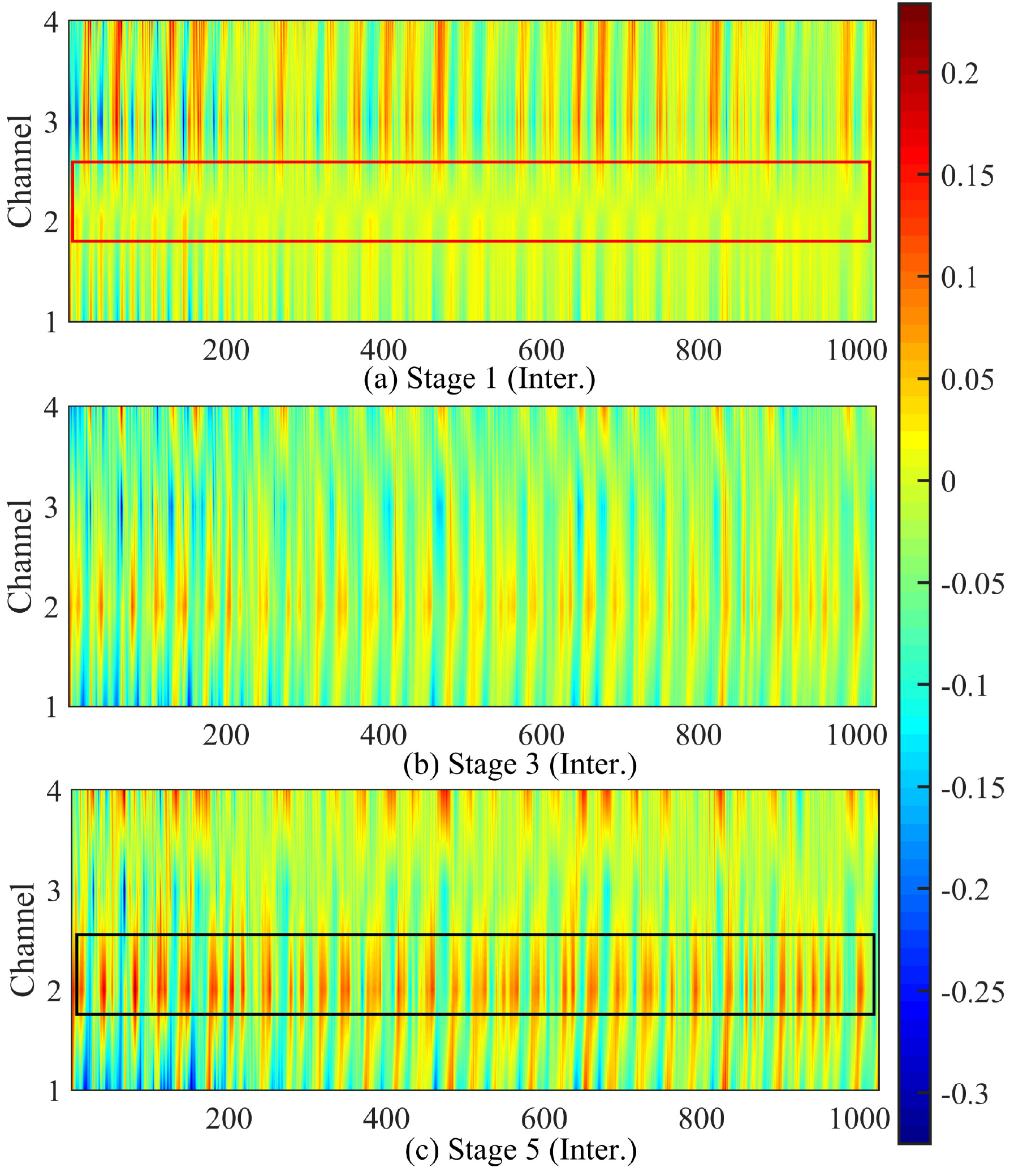}	
	
	\caption{Visualization of hidden state $\mathbf{h}$ within SRNN. The size of $\mathbf{h}$ is $\left(16, 1024 \right)$, where $16$ and $1024$ refer to the channel and feature axis, respectively. We only plot the first 4 channels for convenience. (a) state visualization in the first stage. (b) state visualization in the third stage. (c) visualization in the fifth stage.}
	\label{fig:h-illu}
\end{figure}

\subsection{Insights into feedback learning}
In this subsection, we attempt to analyze the effect of feedback learning. To avoid illustration confusion, we fix the number of stages as 5 herein, i.e., Q $= 5$. First, we give the metric scores in the intermediate stages, and the results are shown in Fig.~{\ref{fig:results-inter}}. One can see that when the first stage is finished, the estimation has similar metric scores over the noisy input in both PESQ and STOI. However, when the network is recursed for more stages, a notable improvement is observed. This indicates that when the estimation from the previous stage is sent back to the network as the feedback component, more prior information can be accumulated and the network is guided to generate cleaner speech estimation. The spectral visualization of the intermediate stages is also presented in Fig.~{\ref{fig:mecha-illu}}. We only give the first, third, and fifth stage herein for convenience. One can see that compared with the input spectrogram, the estimation in the first stage is also relatively noisy. Nevertheless, when more feedback is applied, the noise components are gradually suppressed, which emphasizes the effectiveness of feedback learning. 

As Section~{\ref{SRNN}} states, SRNN is utilized to aggregate the feature information across different stages with a memory mechanism. As such, the hidden state $\mathbf{h}^{l}$ (we omit superscript for simplicity hereafter) is updated in each feedback stage. To emphasize that, we visualize $\mathbf{h}$ in three intermediate stages given Q $= 5$, which is presented in Fig.~{\ref{fig:h-illu}}. As the size of $\mathbf{h}$ is $\left(16, 1024\right)$ (see Table~{\ref{tab1}}), we only extract the first four channels for convenience. One can observe that, for the first stage, SRNN has yet learned clear prior information, leading to blurring feature representation in the hidden state, as shown in Fig.~{\ref{fig:h-illu}} (a), the red box area. When more stages are applied, the SRNN begins to accumulate more prior information about clean speech. As a result, the representation of $\mathbf{h}$ becomes clearer stage by stage, as shown in Fig.~{\ref{fig:h-illu}} (c), the black box area.

\renewcommand\arraystretch{1.0}
\begin{table}[t]
	\caption{The number of trainable parameters among different models. The unit is million. $\textbf{BOLD}$ indicates the lowest trainable parameters.}
	\centering
	\footnotesize
	\begin{tabular}{|c|c|c|c|}
		\hline
		Model &AECNN &RHR-Net &FTNet\\
		\hline
		Para. (million) &6.31 &1.95 &1.02\\
		\hline
	\end{tabular}
	\label{tbl:model-parameters}
	\vspace*{-\baselineskip}
\end{table}

\subsection{Trainable parameters and ideal network depth}  
The number of trainable parameters for the baselines and proposed FTNet is presented in Table~{\ref{tbl:model-parameters}}. One can see that compared with AECNN and RHR-Net, FTNet further decreases the number of trainable parameters, which demonstrates the high parameter efficiency of feedback learning.

To improve network performance, a deeper network is needed, which usually results in more trainable parameters. With feedback learning, the network is reused for multiple stages, and we can explore a deeper network without additional parameters. In this paper, considering the gradient flow, the number of the ideal layers for FTNet is 28$\times$Q, where 28 represents the number of layers for the feedforward gradient flow. Therefore, a deeper network can be explored by recursing the network for more stages.

\section{Conclusions}
\label{sec:conclusions}
In this study, we propose a type of feedback network in the time domain named FTNet for monaural speech enhancement. Stage RNN is proposed to effectively aggregate the deep features across different stages. In addition, concatenated GLUs are adopted to increase the receptive field while controlling the information flow. Experimental results demonstrate that FTNet achieves consistently better performance than the other two advanced time-domain baselines and effectively reduces the number of trainable parameters simultaneously.

\end{document}